\documentclass[twocolumn,showpacs,preprintnumbers]{revtex4}
\usepackage{amssymb}
\usepackage{amsfonts}
\usepackage{amsmath}
\usepackage{graphicx}
\usepackage{dcolumn}
\usepackage{bm}

\input{tcilatex}

\begin{document}

\title{Resistivity peak values at transition between fractional quantum Hall
states}
\author{S. S. Murzin}
\affiliation{Institute of Solid State Physics RAS, 142432, Chernogolovka, Moscow
District, Russia }

\begin{abstract}
Experimental data available in the literature for peak values of the
diagonal resistivity in the transitions between fractional quantum Hall
states ($\rho _{xx}^{\max }$) are compared with the theoretical predictions.
It is found that the majority of the peak values are close to the
theoretical values for two-dimensional systems with moderate mobilities
\end{abstract}

\maketitle

\narrowtext

According to the microscopic\ \cite{Ruz1,Ruz2} and to the scaling \cite{BDD}
theories at transition between two quantum Hall (QH) states\ at low
temperatures the diagonal ($\sigma _{xx}$)\ and Hall ($\sigma _{xy}$)
conductivity components move on the $(\sigma _{xy},\sigma _{xx})$ diagram
along semicircles 
\begin{equation}
\sigma _{xx}^{2}+\left( \sigma _{xy}-\frac{p_{1}/q_{1}+p_{2}/q_{2}}{2}%
\right) ^{2}=\left( \frac{p_{1}/q_{1}-p_{2}/q_{2}}{2}\right) ^{2}  \label{s1}
\end{equation}%
($\sigma _{xx}$ and $\sigma _{xy}$ are in units of $e^{2}/h$). The
semicircles connect pairs of points $(p_{1}/q_{1},0)$ and $(p_{2}/q_{2},0)$,
corresponding to either an integer or fractional quantum Hall state, or an
insulating state ($p_{1}$, $p_{2}$ are the integers and $q_{1}$, $q_{2}$\
are the odd integers). Such semicircles has been observed experimentally 
\cite{Hilke} for the case of transition between quantum Hall state $%
(1/3,\,0) $ and insulating state $(0,\,0)$.

When the magnetic field changes the direct transition between two quantum
Hall states with $\sigma _{xy}=$ $p_{1}/q_{2}$ and $\sigma _{xy}=$ $%
p_{2}/q_{2}$ is allowed at low temperature if and only if \cite{Dolan,Dol2}%
\begin{equation}
|p_{1}q_{2}-p_{2}q_{1}|=1.  \label{rule}
\end{equation}%
Therefore, $\sigma _{xx}$ values in maxima of the transition peaks are equal
to%
\begin{equation}
\sigma _{xx}^{\max }=\frac{\left\vert p_{1}/q_{1}-p_{2}/q_{2}\right\vert }{2}%
=\frac{1}{2q_{1}q_{2}}.  \label{gmax}
\end{equation}%
The semicircle (\ref{s1}) can be replotted on the $(\rho _{xy},\rho _{xx})$
diagram by conventional transformation $\rho _{xy}=\sigma _{xy}/\left(
\sigma _{xy}^{2}+\sigma _{xx}^{2}\right) $ and $\rho _{xx}=\sigma
_{xx}/\left( \sigma _{xy}^{2}+\sigma _{xx}^{2}\right) $. Then all
semicircles (\ref{s1}) but those starting from the point $(0,0)$ transform
into semicircles 
\begin{equation}
\rho _{xx}^{2}+\left( \rho _{xy}-\frac{q_{1}/p_{1}+q_{2}/p_{2}}{2}\right)
^{2}=\left( \frac{q_{1}/p_{1}-q_{2}/p_{2}}{2}\right) ^{2}  \label{Sr}
\end{equation}%
($\rho _{xx}$ and $\rho _{xy}$ are in $h/e^{2}$ units). The semicircles (\ref%
{s1}) starting from the point $(0,0)$ transform into vertical lines at $\rho
_{xy}=1,3,5...$.\ The values of $\rho _{xx}$ at transition peak maxima are
equal to%
\begin{equation}
\rho _{xx}^{\max }=\frac{1}{2p_{1}p_{2}}.  \label{Rmax}
\end{equation}%
\begin{figure}[t]
\includegraphics[width=8.5cm,clip]{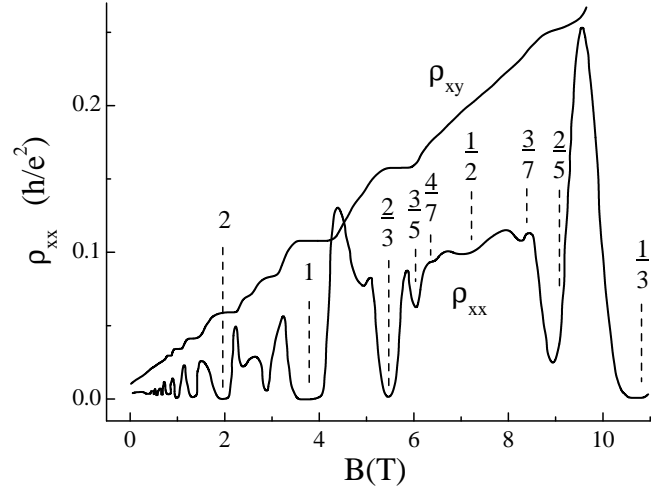}
\caption{Magnetic field dependence of the diagonal ($\protect\rho _{xx}$)
and Hall ($\protect\rho _{xy}$)\ resistivities for $GaAs/Al_{x}Ga_{1-x}As$
heterostructure \protect\cite{Clark 88ss}$^{a}$ with electron density $%
n=9.0\times 10^{10}$ cm$^{-2}$ in a magnetic field perpendicular to the 2D
electron system. Temperature $T=30$ mK. Adopted from Ref.\protect\cite{Clark
88ss}.}
\label{R}
\end{figure}

In the present paper we verify relation (\ref{Rmax}) for appropriate
experimental data available in the literature for different two-dimensional
(2D) systems. 
\begin{table*}[t]
\caption{Values for the current carrier density ($n$ or $p$), the mobility $%
\protect\mu $, and the height of the diagonal resistivity peak $\protect\rho %
_{xx}^{max}$. The references are taken as the numbers of samples. The data
for the different samples or carrier densities from the same paper are
marked additionally by the letters. Data \protect\cite{Tsui 92B2 rho}$%
^{tilt} $ show the resistivity of sample \protect\cite{Tsui 92B2 rho} in
tilted magnetic field.}
\label{T}%
\begin{tabular}{llllllll}
\hline\hline
Transition $\rightarrow $ &  &  & 1/3$\leftrightarrow $2/5 & 2/5$%
\leftrightarrow $3/7 & 1$\leftrightarrow $2/3 & 2/3$\leftrightarrow $3/5 & 
3/5$\leftrightarrow $4/7 \\ 
Value $1/(2p_{1}p_{2})$ $\rightarrow $ &  &  & \textbf{0.25} & \textbf{0.0833%
} & \textbf{0.25} & \textbf{0.0833} & \textbf{0.0417} \\ 
&  &  &  &  &  &  &  \\ 
Sample & Density & Mobility &  &  & $\rho _{xx}^{\max }$ &  &  \\ 
& (10$^{10}$cm$^{-2}$) & (cm$^{-2}/Vs$) &  &  &  &  &  \\ 
\tableline GaAs/AlGaAs \cite{Clark 88ss}$^{a}$ & n=9.0 & $\approx $1$\times $%
10$^{6}$ & 0.25 &  &  &  &  \\ 
GaAs/AlGaAs \cite{Clark 88ss}$^{b}$ & n=15 & $\approx $1$\times $10$^{6}$ & 
&  &  & 0.085 &  \\ 
GaAs/AlGaAs \cite{Clark 88ss}$^{c}$ & n=18 & 1.9$\times $10$^{6}$ &  &  &  & 
0.074 &  \\ 
GaAs/AlGaAs \cite{Clark 88ss}$^{d}$ & n=19 & $\approx $1$\times $10$^{6}$ & 
&  & 0.126 &  &  \\ 
GaAs/AlGaAs\ \cite{Clark 88L} & n=9.5 & 1$\times $10$^{6}$ &  &  &  &  & 
0.042 \\ 
GaAs/AlGaAs \cite{Koch} & n=21 & 2.2$\times $10$^{5}$ &  &  &  & 0.068 &  \\ 
GaAs/AlGaAs \cite{Tsui 93L reentrant} & n=6.6 & 1.4$\times $10$^{6}$ &  & 
0.075 &  &  & 0.056 \\ 
GaAs/AlGaAs \cite{Tsui 92B2 rho} & n=2.4 & 7$\times $10$^{5}$ & 0.26 &  &  & 
0.117 &  \\ 
GaAs/AlGaAs \cite{Tsui 92B2 rho}$^{tilt}$ & n=2.4 & 7$\times $10$^{5}$ & 0.25
&  &  &  &  \\ 
GaAs/AlGaAs \cite{Kang} & n=35 &  &  &  &  & 0.077 &  \\ 
GaAs/AlGaAs \cite{Tsui 96 Scince} & n=6.5 & 5.5$\times $10$^{5}$ & 0.33 &  & 
0.25 &  &  \\ 
SiGe/Si/SiGe \cite{Lai 04B FQHE} & n=45 & 7.5$\times $10$^{4}$ &  &  & 0.22
&  &  \\ 
SiGe/Si/SiGe \cite{Lai 04L Two} & n=27 & 2.5$\times $10$^{5}$ & 0.22 &  &  & 
&  \\ 
GaAs quantum well \cite{Mills 99L hole} & p=8.7 & $>5\times 10^{5}$ &  &  & 
0.27 &  &  \\ 
GaAs quantum well \cite{Tsui 05L holes}$^{a}$ & p=1.2 &  &  &  & 0.24 &  & 
\\ 
GaAs quantum well \cite{Tsui 05L holes}$^{b}$ & p=1.5 &  &  &  & 0.22 &  & 
\\ 
GaAs/AlGaAs \cite{Pepper 95B hole} & p=12 & 5.4$\times $10$^{5}$ &  &  &  & 
0.077 &  \\ 
GaAs/AlGaAs \cite{Rodgers 93cm2 hole} & p=16 & 1.5$\times $10$^{5}$ & 0.41 & 
& 0.35 & 0.13 &  \\ 
GaAs quantum well \cite{Shayegan 94B hall} & p=15 & 1$\times $10$^{6}$ &  & 
0.18 & 0.29 & 0.19 & 0.15 \\ \hline\hline
\end{tabular}%
\end{table*}
The theories \cite{Ruz1,Ruz2,BDD} have been developed for spinless (or
totally spin polarized) electrons. Therefore, we discuss only transitions $%
p_{1}/q_{1}\leftrightarrow p_{2}/q_{2}$ with $p_{1}/q_{1}$ and $%
p_{2}/q_{2}\leq 1$. For short $p_{1}/q_{1}\leftrightarrow p_{2}/q_{2}$
denote the transition between QH states with $\sigma _{xy}=p_{1}/q_{1}$ and $%
p_{2}/q_{2}$. Note that spin depolarization of 2D electron system in
GaAs/AlGaAs structures is possible even for $p_{1}/q_{1}<1$. Good agreement
of the experimental data\ with the theoretical results is found only for
samples with mobility $\mu \lesssim 2\times 10^{6}$ cm$^{-2}/Vs$, and so,
here we consider only such samples. The references are taken as the numbers
of samples. The data for different samples or carrier densities from the
same paper are marked additionally by the letters.

In Fig.\ref{R} we show example (sample no.\cite{Clark 88ss}$^{a}$) of the
diagonal ($\rho _{xx}$) and Hall ($\rho _{xy}$) resistivities as a function
of the magnetic field for the modulation-doped heterostructure $%
GaAs/Al_{x}Ga_{1-x}As$ with electron density $n=9\times 10^{10}$~cm$^{-2}$
and mobility ($\mu $) about 10$^{6}$~cm$^{2}$/Vs.\ In the figure there is
only one sufficiently clear defined peak at transition $2/5\leftrightarrow
1/3$ with $1/(2p_{1}p_{2})=0.25$ and the same experimental value of $\rho
_{xx}^{\max }=0.25$. Experimental values of $\rho _{xx}^{\max }$ which we
have found in the literature are listed in Table I. These data are chosen
because they satisfy the following conditions: (i) the temperature is low, $%
T<100$ mK; (ii) the transitions are pronounced ($\rho _{xx}$ in both
neighboring minima is smaller than $0.3\rho _{xx}^{\max }$), (iii) the
transitions are sufficiently narrow (width of the peak at the level $%
2/3(\rho _{xx}^{\max })$ is smaller than \ $\left\vert B_{\max
}-B_{1}\right\vert $ and $\left\vert B_{\max }-B_{2}\right\vert $, where $%
B_{\max }$ is the magnetic field in the maximum, $B_{1}$ and $B_{2}$ are
the\ magnetic fields in the neighboring minima). Experimental data are
disregarded if the current through the sample was rather large ($>20$ nA),
edge effects were knowingly\ essential \cite{Gol,Gol2}

For $GaAs/AlGaAs$ heterostructures the majority of the experimental values
of $\rho _{xx}^{\max }$ are in agreement with the theoretical predictions $%
1/(2p_{1}p_{2})$ with accuracy about $10\%$. Only the values of $\rho
_{xx}^{\max }$ at transition $1\leftrightarrow 2/3$ in sample \cite{Clark
88ss}$^{d}$, at transition $2/3\leftrightarrow 3/5$ in sample \cite{Tsui
92B2 rho}, at transition $3/5\leftrightarrow 4/7$ in sample \cite{Tsui 93L
reentrant}, and at transition $2/5\leftrightarrow 1/3$ in sample \cite{Tsui
96 Scince} significantly differ from $1/(2p_{1}p_{2})$. The differences at
transitions $1\leftrightarrow 2/3$\ and $2/3\leftrightarrow 3/5$ are due to
spin depolarization. This is confirmed \cite{Tsui 92B2 rho} by intense
changing of the curve $\rho _{xx}(B)$ in tilted magnetic field \cite{Tsui
92B2 rho}$^{tilt}$ up to the field where $p_{1}/q_{1}=4/7$. The difference $%
\rho _{xx}^{\max }$ from $1/(2p_{1}p_{2})$\ at transitions $%
3/5\leftrightarrow 4/7$ in sample \cite{Tsui 93L reentrant} is probably also
defined by spin depolarization. Note, that at transition $3/5\leftrightarrow
4/7$ in sample \cite{Clark 88L} $\rho _{xx}^{\max }=0.042$ is nearly $%
1/(2p_{1}p_{2})=0.0417$ and at transition $2/3\leftrightarrow 3/5$ the
values of\ $\rho _{xx}^{\max }\approx 1/(2p_{1}p_{2})=0.0833$ in four
samples \cite{Clark 88ss}$^{b}$, \cite{Clark 88ss}$^{c}$, \cite{Koch}, and 
\cite{Kang}. Probably in these samples 2D electron systems are totally
polarized in the regions of transition $3/5\leftrightarrow 4/7$ and $%
2/3\leftrightarrow 3/5$.

In two n-type $Si_{1-x}Ge_{x}/Si/Si_{1-x}Ge$ quantum wells experimental
values of $\rho _{xx}^{\max }$ are close to $1/(2p_{1}p_{2})$ at transitions 
$1\leftrightarrow 2/3$ observed in sample \cite{Lai 04B FQHE} and at
transition $1/3\leftrightarrow 2/5$ observed in sample \cite{Lai 04L Two}.

In p-type GaAs quantum wells \cite{Mills 99L hole} and \cite{Tsui 05L holes}
with two different hole densities \cite{Tsui 05L holes}$^{a}$ and \cite{Tsui
05L holes}$^{b}$ $\rho _{xx}^{\max }$ values are close to the value of $%
1/(2p_{1}p_{2})=0.25$ at transitions $1\leftrightarrow 2/3$. Besides $\rho
_{xx}^{\max }=0.077\approx 1/(2p_{1}p_{2})=0.0833$ for 2D\ hole system \cite%
{Pepper 95B hole} in $GaAs/Al_{x}Ga_{1-x}As$ heterostructure. For two $p$%
-type samples \cite{Rodgers 93cm2 hole} and \cite{Shayegan 94B hall} $\rho
_{xx}^{\max }$ values significantly differ from $1/(2p_{1}p_{2})$, probably
because the temperatures were not sufficiently low.

Disagreement between experimental and theoretical results for samples with
mobility $\mu \gtrsim 2\times 10^{6}$ cm$^{-2}/Vs$ can be caused by
insufficiently low temperature, by edge effects, or by weak macroscopic
inhomogeneities of the sample \cite{Rusin3}.

In summary, at low temperature the peak values of the diagonal resistivity
at the transitions between fractional quantum Hall states ($\rho _{xx}^{\max
}$) are found to be close to the theoretically predicted values $%
1/(2p_{1}p_{2})$ for majority of two-dimensional systems with moderate
mobilities.

This work was supported by \emph{Russian Foundation for Basic Research.}

\end{document}